\documentclass{article}

\usepackage{arxiv}

\usepackage[utf8]{inputenc} 
\usepackage[T1]{fontenc}    
\usepackage{hyperref}       
\usepackage{url}            
\usepackage{booktabs}       
\usepackage{amsfonts}       
\usepackage{nicefrac}       
\usepackage{microtype}      
\usepackage{lipsum}		
\usepackage{graphicx}
\usepackage{natbib}
\usepackage{doi}
\usepackage{wrapfig}

\usepackage{lscape}
\usepackage{longtable}
\usepackage{hyperref}
\usepackage{xltabular}
\usepackage{caption}
\usepackage{tabularray}
\usepackage{tablefootnote}
\usepackage{enumerate}
\usepackage{array}

\title{A Survey on Food Ingredient Substitutions}

\author{\href{https://orcid.org/0009-0005-6172-0129}{\includegraphics[scale=0.06]{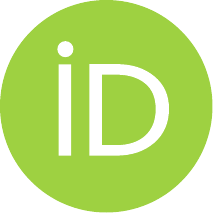}\hspace{1mm}Hyunwook ~Kim} \\
	AI Institute\\
	University of South Carolina\\
	Columbia, South Carolina\\
	\texttt{hyunwook@email.sc.edu} \\
	\And
	\href{https://orcid.org/0000-0002-5642-3438}{\includegraphics[scale=0.06]{orcid.pdf}\hspace{1mm}Revathy ~Venkataramanan} \\
	AI Institute\\
	University of South Carolina\\
	Columbia, South Carolina\\
	\texttt{revathycv24@gmail.com} \\
 \And
	\href{https://orcid.org/0000-0002-0021-5293}{\includegraphics[scale=0.06]{orcid.pdf}\hspace{1mm}Amit ~Sheth} \\
	AI Institute\\
	University of South Carolina\\
	Columbia, South Carolina\\
	\texttt{amit@sc.edu} \\
}

\renewcommand{\shorttitle}{A Survey on Food Ingredient Substitutions}

\hypersetup{
pdftitle={A template for the arxiv style},
pdfsubject={q-bio.NC, q-bio.QM},
pdfauthor={David S.~Hippocampus, Elias D.~Striatum},
pdfkeywords={First keyword, Second keyword, More},
}

\begin{document}
\maketitle

\begin{abstract} Diet plays a crucial role in managing chronic conditions and overall well-being. As people become more selective about their food choices, finding recipes that meet dietary needs is important. Ingredient substitution is key to adapting recipes for dietary restrictions, allergies, and availability constraints. However, identifying suitable substitutions is challenging as it requires analyzing the flavor, functionality, and health suitability of ingredients. With the advancement of AI, researchers have explored computational approaches to address ingredient substitution. This survey paper provides a comprehensive overview of the research in this area, focusing on five key aspects: (i) datasets and data sources used to support ingredient substitution research; (ii) techniques and approaches applied to solve substitution problems  (iii) contextual information of ingredients considered, such as nutritional content, flavor, and pairing potential; (iv) applications for which substitution models have been developed, including dietary restrictions, constraints, and missing ingredients; (v) safety and transparency of substitution models, focusing on user trust and health concerns. The survey also highlights promising directions for future research, such as integrating neuro-symbolic techniques for deep learning and utilizing knowledge graphs for improved reasoning, aiming to guide advancements in food computation and ingredient substitution. \end{abstract}

\keywords{Ingredient substitution \and Food Computation \and Ingredient Knowledge Graphs \and Computational food science \and Food ingredient substitution models \and Recipe adaptation techniques}

\section{Introduction}
Food computing is an interdisciplinary field that applies computational approaches to address challenges and opportunities within the food domain. It involves acquiring and analyzing heterogeneous food data from disparate sources for perception, recognition, retrieval, recommendation, prediction, and monitoring of food (\cite{min2019survey}). Perception in food computing refers to the ability of systems to interpret sensory data, such as visual or olfactory information, to understand food characteristics and user preferences. Recognition involves using computer vision techniques to accurately identify and classify food items, enhancing applications such as dietary tracking and food safety monitoring. Retrieval encompasses methods for accessing food-related information from databases, allowing users to quickly find specific recipes, nutritional data, or ingredient substitutes. Recommendation systems leverage user data and preferences to suggest recipes, meal plans, or food products that align with individual dietary needs and taste profiles. Finally, monitoring involves tracking food consumption, quality, and freshness through various sensors and technologies, enabling better food management and waste reduction. Together, these components create a robust framework that enhances our understanding and interaction with food, ultimately leading to improved health outcomes and culinary experiences.

The focus of this survey is food ingredient substitutions, a specialization that aims to develop algorithms and systems to suggest alternative ingredients in recipes, addressing dietary restrictions, allergies, or availability constraints. This study is not limited to one subcategory presented above but combines multiple subcategories, such as recommendation and recognition. By leveraging computational methods, researchers in this field seek to enhance and personalize culinary experiences while accommodating individual needs and preferences.

Recent research underscores the profound influence of food on our overall well-being and the management of chronic conditions. In fact, the gut is referred to as the “second brain” (\cite{harvard_brain_gut_2024}) by several studies, opening new perspectives on the depth of its impact on our health and well-being, as well as potential new treatment options (\cite{ochoa2016second}). Due to increasing awareness on the importance of healthier eating habits, people have become more cautious about their food choices. Studies have shown a significant increase in the healthy eating habit trend among individuals (\cite{fang2023global}). With the rise of diverse diet options, such as keto or Paleo, and dietary restrictions, such as lactose intolerance or nut allergies, there is a need for personalization in recipes that suit individual needs. Adapting recipes to accommodate specific health needs, such as managing diabetes or adhering to a gluten-free diet, requires careful consideration of ingredient choices and cooking methods.

The choice of ingredients plays a pivotal role in personalizing recipes to suit an individual's health condition, food preferences, and nutritional needs compared to other factors involved in a recipe. Ingredients serve as the building blocks of a dish, determining its flavor, texture, and nutritional composition. Unlike other factors involved in a recipe, the choice of ingredients directly impacts the nutritional content and health implications of the dish. For individuals managing chronic conditions like diabetes, heart disease, or food allergies, selecting ingredients that are low in sugar, sodium, or allergens is essential for maintaining health and well-being. Similarly, those following specific dietary patterns, such as vegetarian, vegan, or gluten-free diets, rely on ingredient selection to ensure compliance with their dietary restrictions. Furthermore, ingredients also play a crucial role in enhancing the flavor profile of a recipe. The choice of ingredients stands out as a critical factor in recipe personalization, offering a versatile and effective means of tailoring meals to meet individual health needs, food preferences, and nutrition goals.

Finding alternative cooking methods is relatively effortless as the nutrition majorly affected by the mediums used for cooking such as fat, liquid, or dry heat. Whereas identifying ingredient substitution is challenging as the substitution not only needs to satisfy the dietary needs and flavor preferences, but also the functional role of the original ingredient in the recipe. In the case of a cake, eggs act as a binding agent, and finding an appropriate vegan substitute is crucial to maintain the desired texture and structure of the final product. On the other hand, in a dish like scrambled eggs, the functionality of the eggs is different, and a substitute like tofu may be more suitable. These substitutes must also maintain the flavor and texture of the original recipe. That is, the choice of substitution is subjective to a recipe and needs to be carefully considered in the context of the specific dish in addition to user’s preferences.

There have been several research studies over the time that employ food computations techniques for ingredient substitution. They specifically leverage the data-driven approaches such as deep learning and machine learning to analyze suitability of ingredients in a given context. Given the complexity of identifying appropriate ingredient substitution to personalize a recipe, these AI models can aid individuals to cater a recipe based on their health condition and food preferences. This not only empowers individuals to make informed choices about their diet but also enhances their ability to adhere to dietary recommendations and achieve better health outcomes.

In addition to recipe personalization for an individual’s needs, ingredient substitution models can aid the culinary domain by creating new explorative areas. By leveraging vast databases of recipes and ingredient interactions, AI models can suggest innovative ingredient substitutions that chefs may not have considered. This fosters creativity in recipe development and encourages experimentation with new flavor combinations. It can recommend lesser-known or underutilized ingredients as substitutes, encouraging exploration and discovery of new flavors and ingredients. These AI models can comprehensively consider several factors about a given ingredient such as nutrition, behavior, functionality with respect to recipe and so on to analyze the suitability of substitutions. Suggesting alternative ingredients in case of missing ingredients can also aid the chefs in creating recipes with swapped ingredients. Overall, these AI models for ingredient substitutions can open up new possibilities in the culinary domain in generating creative variations, adapting to dietary preferences, cuisine fusions and recipe customizations.

In this survey, we focus on investigating research studies on ingredient substitution on various factors such as creation of datasets, types of techniques implemented, curation of context and knowledge and safety of the system in terms of transparency and trustworthiness. Unlike datasets for recipes, there are limited sources and documentation that provide data on ingredient substitution. In particular, finding ingredient substitution dataset in the context of recipes is a rarity. Several works either used the dataset from existing sources or created datasets based on large recipe databases. To add, ingredients need to be analyzed in multiple contextual knowledge such as nutrition, dietary, disease, and flavor context. A few of the research studies have attempted to utilize one or many of these contextual knowledge of ingredients. We also aim to inspect the types of AI methods that have been utilized to solve ingredient substitution. To add, based on the AI methods employed and data sources used, we evaluate and report the safety of these systems in terms of transparency and trustworthiness. Specifically, we aim to answer the following research questions:

\textbf{RQ1}: What are the commonly used data sources and techniques for ingredient substitution dataset creation?

\textbf{RQ2}: What are the technological approaches proposed to address challenges in ingredient substitution?

\textbf{RQ3}: What are the additional contextual factors of ingredients that are considered while recommending a substitute ingredient for a given ingredient?

\textbf{RQ4}: What are the dietary and health constraints for which ingredient substitution techniques have been developed?

\textbf{RQ5}: How safe are these models or systems in terms of transparency and trustworthiness?

To the best of our knowledge this is the first survey on food ingredient substitution that uses AI-based food computation approaches to analyze the suitability of ingredients. The goal of this work is to highlight the research gap for future directions and study the existing datasets and methods proposed to solve the ingredient substitution problem. The specific contributions of the paper are as follows (i) exploring the sources and methodologies employed in generating datasets for ingredient substitution, (ii) survey the models and techniques utilized in developing ingredient substitution systems, (iii) review the contextual knowledge of ingredients and the dietary restrictions considered while recommending ingredient substitution, and (iv) presenting the discoveries, identifying research gaps, and outlining future directions in the domain of ingredient substitution.

\section{Background}
\subsection{Ingredient Substitution} 
Food ingredient substitution refers to the task of replacing one ingredient with another in a recipe while aiming to maintain the intended flavor, texture, and overall recipe experience. This can be necessitated by various factors such as \textit{dietary restrictions}, \textit{dietary constraints}, \textit{health conditions}, or dealing with ingredient shortages. \textit{Dietary restrictions} are conditions necessitated by health-related issues, such as lactose intolerance, allergies, or celiac disease, which requires a gluten-free diet. Conversely, \textit{dietary constraints} encompass personal dietary choices influenced by cultural, religious, or lifestyle factors, such as veganism, vegetarianism, or ketogenic diets. \textit{Health conditions} include chronic conditions such as diabetes or hypertension that require dietary modifications to maintain the chronic condition. The process involves selecting alternative ingredients that possess similar functional properties or flavor profiles to the original ingredient, ensuring that the dish remains cohesive and satisfying.

\subsection{Ingredient Substitution Model}
While there is not a single universally accepted definition of an ingredient substitution model, this survey paper defines ingredient substitution models as AI-driven systems that analyze and recommend suitable replacements for ingredients in recipes. These systems may take into account factors such as dietary restrictions, health conditions, flavor profiles, ingredient functionality, and user preferences.

\subsection{Food Recognition}
While previous survey papers such as \cite{min2019survey} do not directly focus on food ingredient substitutions, it is worth illuminating previous research in terms of food ingredient substitutions to aid future research in the field. For example, a considerable body of research on food perception is predominantly situated within the domains of neuroscience and cognitive science (\cite{killgore2003cortical, sorensen2003effect, rosenbaum2008leptin, medic2016presence}). Accurate ingredient detection allows consumers to make informed choices based on their dietary needs, allergies, or ethical preferences. In a similar vein, food recognition assumes a pivotal role, especially ingredient recognition, in addressing concerns related to health and dietary considerations, including conditions like diabetes and cardiovascular diseases. A prominent method of food recognition involves the utilization of food images, and a considerable body of research has been dedicated to proposing food recognition algorithms aimed at addressing challenges associated with health-focused decision-making (\cite{hoashi2010image, martinel2018wide, martinel2015structured, yang2010food, kagaya2014food, wu2016learning}). More specifically, some research focuses on enhancing health and dietary management through computational tools like eDental and FoodSG (\cite{zheng2022edental, zheng2023plate, goh2024development}). The eDental tool uses food recognition techniques to improve dental care by tracking dietary intake, while FoodSG supports healthcare applications in Singapore by identifying food consumption patterns and providing medical-grade nutrient information. Their work includes the release of a localized food dataset, FoodSG-233, to aid in these efforts.

\subsection{Food Retrieval}
Furthermore, retrieval methodologies can be utilized to identify ingredient-substitution pairs from various types of datasets. Retrieval methods in food computing involve the construction of datasets containing recipes, food images, and ingredient information from various sources. This compiled data can enable cross-modal retrieval of suitable ingredient substitutes based on visual or textual queries (\cite{kitamura2009foodlog, aizawa2014comparative, chen2017cross}).

\subsection{Food Recommendation}
Recommendation systems in food computing can also be used to suggest ingredient substitutions tailored to individual needs, such as health issues, cultural preferences, or context-driven factors. Recommending viable substitutes empowers consumers to make healthier or more sustainable choices without compromising taste or texture. While most research on food recommendations has focused on recipe-oriented recommendations, the scope of recommendations can be expanded to include ingredient recommendations. This research can vary widely, encompassing context-driven recommendations based on factors such as gender, time, or cultural aspects (\cite{cheng2017influence}), as well as recommendations related to health issues (\cite{rokicki2018impact, nag2017live}). Additionally, this research leverages multi-modal data and external knowledge to address constraints such as allergen avoidance (\cite{pallagani2022rich}).

\subsection{Open-source Data for Ingredient Substitution}
Additionally, prediction and monitoring techniques can leverage large-scale food dataset creation, especially for ingredient substitution research, where datasets are lacking. This data can be sourced from websites and social media to gain insights into public health trends, popular recipes, and dietary patterns. Such insights can inform strategies for promoting healthier ingredient substitutions or identifying critical ingredients for food security analysis. Given the substantial volume of data available from websites and social networks, the domain of food-related prediction and monitoring has the capacity to accumulate diverse health- and food-related insights. This includes the collection of statistical data, such as diabetes statistics (\cite{abbar2015you}), and the curation of popular recipes sourced from social media platforms (\cite{sanjo2017recipe}), as well as the potential for public health monitoring (\cite{capurro2014use}).

\subsection{Significance of Ingredient Substitution}
\begin{figure}[!ht]
  \begin{center} 
    \includegraphics[width = 0.6\textwidth] {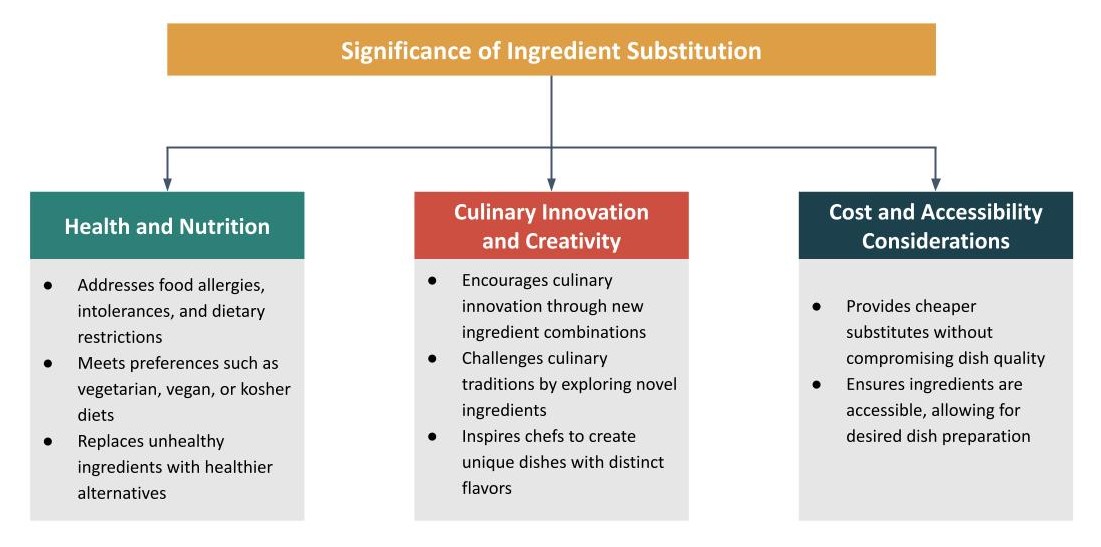}
    \caption{Significance of Ingredient Substitution in the areas of Health, Culinary Innovations and Accessibility Considerations}
    \label{fig:significance}
  \end{center}
\end{figure}

This section presents the significance of ingredient substitution categorized into three major areas as shown in Figure \ref{fig:significance}
\subsubsection{Health and Nutrition} 
\begin{itemize}
\item Dietary Restrictions: Many individuals have food allergies or intolerances that require them to avoid certain ingredients. Studying ingredient substitution provides insights into suitable alternatives that can be safely incorporated into recipes without compromising taste or nutritional value. For instance, substituting wheat flour with gluten-free flour allows individuals with celiac disease or gluten sensitivity to enjoy a variety of baked goods.

\item Dietary Constraint: Ingredient substitution also enables individuals to follow specific dietary preferences, such as vegetarian, vegan, or kosher diets, to adhere to their restrictions while still enjoying flavorful and satisfying meals. For example, substituting animal-based ingredients with plant-based alternatives enables individuals following vegetarian, vegan, or kosher diets to adhere to their dietary restrictions while still enjoying flavorful and satisfying meals.

\item Health Conditions: Diet plays a major role in the prevention and management of chronic health conditions, such as diabetes and hypertension. Appropriate ingredient substitutions that focus on low-carb or low-sodium would be highly beneficial to individuals with health conditions.

\item Nutritional Improvements: Ingredient substitution can also facilitate the creation of more nutritious dishes by replacing high-fat, high-calorie, or high-sugar ingredients with healthier alternatives. This can contribute to overall health and well-being by reducing the intake of certain nutrients that may be detrimental in excess.
\end{itemize}

\subsubsection{Culinary Innovation and Creativity}
\begin{itemize} 
\item Creating new recipes: Exploring ingredient substitution encourages culinary innovation and creativity by challenging traditional recipes and techniques. Chefs and home cooks can experiment with novel ingredients and flavor combinations to create unique and innovative dishes. Studying ingredient substitution helps uncover new possibilities for culinary expression and expands the repertoire of flavors and textures available to cooks.
\end{itemize}

\subsubsection{Cost and Accessibility Considerations}
\begin{itemize}
\item Ingredient substitution also addresses considerations related to food cost and ingredient availability. When certain ingredients are unavailable or unaffordable, substitutions can provide cost-effective and accessible alternatives, allowing cooks to prepare their desired dishes without compromising the overall quality and flavor.
\end{itemize}

\begin{figure}[!ht]
  \begin{center} 
    \includegraphics[width = 0.6\textwidth]{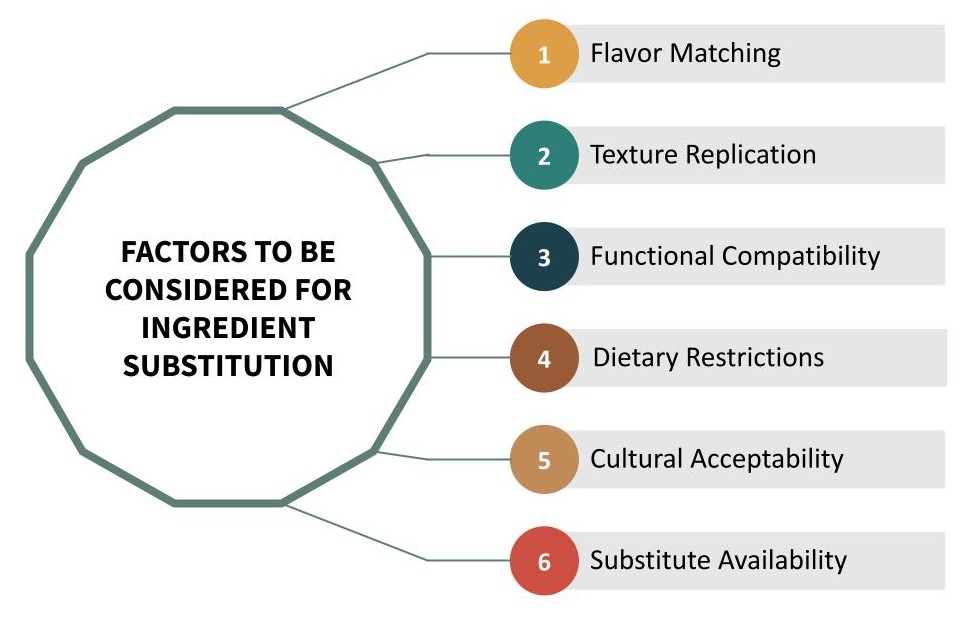}
    \caption{Key Factors for Ingredient Substitution: Ensuring Flavor, Texture, Functionality, and Suitability in Culinary Decisions}
    \label{fig:factors}
  \end{center}
\end{figure}

\subsection{Challenges}
While ingredient substitution offers numerous benefits, it presents challenges that chefs and home cooks must navigate. The factors influencing ingredient substitutions include flavor matching, texture replication, and functional compatibility with the original ingredient as shown in Figure \ref{fig:factors}. Additionally, considerations such as dietary restrictions, cultural acceptability, and the availability of substitutes further shape the decision-making process for chefs and home cooks. One significant challenge is ensuring that the substitute ingredient effectively replicates the flavor, texture, and functionality of the original ingredient. Additionally, individuals with dietary restrictions or allergies may encounter difficulty finding suitable substitutes that meet their specific dietary needs without compromising taste or nutritional value. Moreover, cultural preferences and culinary traditions may influence the acceptability of certain substitutes, further complicating the process of ingredient substitution. It is well established that ingredient substitutions play a crucial role in the culinary arts and food computing domains, catering to the diverse needs and preferences of users. These substitutions can address a wide range of considerations, from dietary restrictions and health concerns to ingredient availability and cost. Despite these challenges, advancements in food computing approaches offer promising solutions to enhance the efficiency and accuracy of ingredient substitution, thereby addressing the evolving needs of modern cuisine.

\section{Survey Approach}
\subsection{Scope of the survey}
The scope of the survey is to investigate and present the datasets and methods along with additional contextual knowledge employed for food ingredient substitution. It identifies commonly utilized data sources and techniques for creating ingredient substitution datasets and delves into various technological approaches to address inherent challenges. The paper examines factors considered when recommending substitute ingredients and explores dietary and health constraints focused on by substitution techniques. Additionally, it assesses the safety, transparency, and trustworthiness of these models or systems. Through this analysis, the paper provides a comprehensive overview of current ingredient substitution research, highlighting advancements, challenges, and future directions.

\subsection{Databases Indexed}
The articles considered for this survey were gathered from a diverse range of databases and publication venues spanning computer science, natural language processing, food science, and multidisciplinary domains. In the field of computing, the Association for Computing Machinery (ACM) and the Institute of Electrical and Electronics Engineers (IEEE) digital libraries were indexed, including proceedings from conferences such as the ACM International Joint Conference on Pervasive and Ubiquitous Computing and the IEEE Pacific Rim Conference on Communications, Computers and Signal Processing (PACRIM). For natural language processing, the proceedings of the Conference of the Asia-Pacific Chapter of the Association for Computational Linguistics and the International Joint Conference on Natural Language Processing (ACL-IJCNLP) were reviewed. The Journal of Food Composition and Analysis, a leading venue in food science and analysis, was also indexed. Additionally, multidisciplinary journals like Frontiers in Artificial Intelligence, Procedia Computer Science, and the ACM Multimedia Workshop on Multimedia for Cooking and Eating Activities were considered to capture relevant research from diverse fields. 

\subsection{Search Terms}
To identify the most relevant literature for this survey on food ingredient substitution, we employed a two-pronged search strategy. First, we conducted a broad keyword search on Google Scholar and other databases mentioned in section 3.2 using the terms “ingredient substitution” and “food ingredient.” This initial search allowed us to uncover highly cited and influential works in the field, such as the paper titled “Identifying ingredient substitutions using a knowledge graph of food” (\cite{shirai2021identifying}).

Subsequently, we employed a snowballing technique, carefully reviewing the references cited in the key articles identified through the initial keyword search. This approach enabled us to expand the scope of our survey and uncover additional relevant research papers that may not have been directly retrieved through the keyword search. Through this combined strategy of targeted keyword searches and reference snowballing, we were able to assemble a comprehensive collection of 28 research articles that formed the basis for our analysis and survey on food ingredient substitution.

\subsection{Inclusion and Exclusion Criteria}
We contain our scope to food ingredient substitution survey to research works that utilize food computation methods to address the challenges. There have been studies in the Food Science journal database that examine the characteristics of a particular recipe with specific ingredient substitutions. The goal of these studies is to analyze if a particular ingredient can be substituted in a particular recipe while retaining the flavor of the recipes (\cite{iguttia2011substitution, guler2021novel}). As the scope of our survey paper is to investigate the data-driven approaches such as machine learning and deep learning algorithms utilized to solve ingredient substitution, the papers focusing on studying the characteristics of a specific dish with substituted ingredients are excluded. Additionally, while studies in the food science domain that aim to identify replaceable food ingredients based on the comparison of bioactivity patterns offer valuable insights, they do not align with the specific objective of our research. Therefore, research such as the NaturaPredictaTM model (\cite{kim2023naturapredicta}), which utilizes NLP techniques to predict bioactivity and similarity to existing approved ingredients for the development of new health functional food ingredients, is excluded from our survey. Our aim is to provide a comprehensive analysis of the advancements specifically in data-driven approaches for recipe ingredient substitution, ensuring a focused and meaningful exploration of the subject matter.

\section{Overview of the collected papers}
\begin{figure}[!ht]
  \begin{center} 
    \includegraphics[width = 0.7\textwidth]{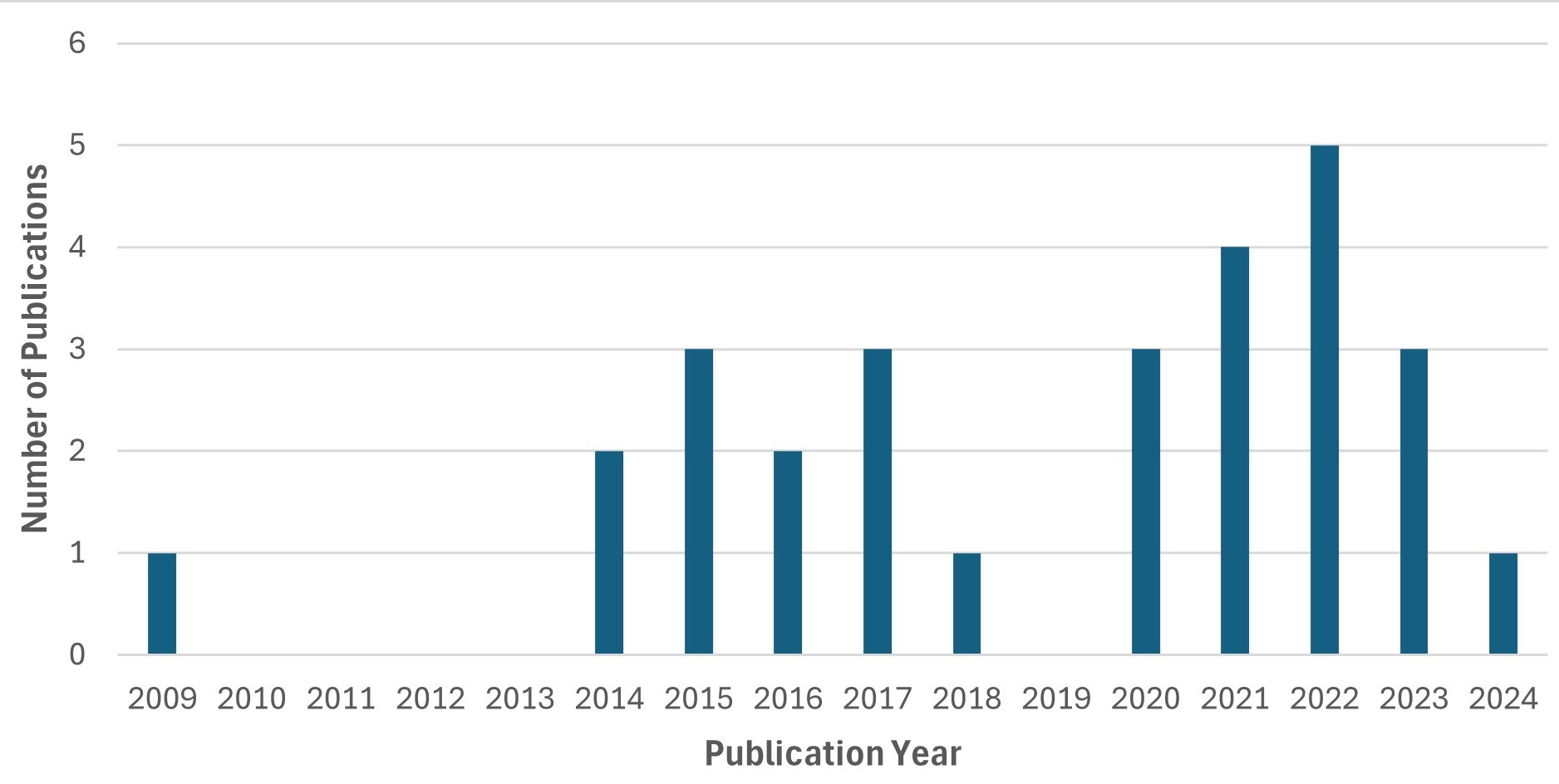}
    \caption{The Number of Papers Published from 2009 to 2024}
  \label{fig:evaluation}
  \end{center} 
\end{figure}

Figure \ref{fig:evaluation} illustrates the frequency of publications for various research papers across the years from 2009 to 2024. In 2014, there was a notable increase with two publications, attributed to studies by \cite{boscarino2014automatic} and \cite{saengsupawat2014ontology}. The trend continues to rise in 2015 with three publications, referencing research by \cite{ooi2015ingredient}, \cite{yamanishi2015alternative}, and \cite{gaillard2015improving, gaillard2017adaptation}. The trend continued to rise from 2020 to 2022. Particularly, in 2022, there was a significant increase to five publications, with contributions from \cite{pacifico2022automatic, pacifico2022improved}, \cite{li2022food}, \cite{loesch2022automated}, and \cite{lawrynowicz2022food}. In 2023, the number of publications dropped slightly to three, such as \cite{antognini2022assistive}, \cite{fatemi2023learning}, and \cite{bikakis2023graphical}. The most recent data available for 2024 shows one publication, attributed to \cite{loesch2024automated}. In total, there are 28 papers published from 2009 to 2024 on ingredient substitutions.

\renewcommand{\arraystretch}{1.5}
\begin{longtable}{
  p{0.3\linewidth} p{0.6\linewidth}}
\caption{Comprehensive Overview of Existing Works: Ingredient Substitutions through 4 Key Questions—Who, What, Why, and How}
\label{tab:overview}\\

\hline
\textbf{Dimension} & \textbf{Examples} \\
\hline
Who are the users? & - Individuals who have diet restrictions, diet constraints, or health conditions \newline - People with allergies to specific ingredients \newline - Individuals seeking healthier food alternatives \newline - Those requiring meal options when ingredients are scarce or unavailable \newline - Individuals interested in experiencing new local culinary styles\\ \hline
What is being recommended? & - Alternative or substitute ingredient: For example, you can substitute butter with sour cream for reduced fat and calories\\ 
\hline
Why is it being recommended? & - Nutrition values \newline - Users' dietary restrictions such as allergens, lactose-free or gluten-free \newline - Users who have dietary constraints such as vegan or vegetarian \newline - Users who have health conditions such as diabetes, hypertension and so on \newline - Missing ingredients\\ 
\hline
How are ingredient substitutions generated? & - Rule-based approaches \newline - Systems that utilize knowledge graphs to generate substitutions \newline - Using learnable models such as machine or deep learning models\\
\hline
\end{longtable}

Table \ref{tab:overview} systematically categorizes ingredient substitution research by target users, recommended items, and the underlying motivation for the suggestions. By organizing this information, the table offers a comprehensive overview of the varied methodologies and objectives driving food recommendation research. Specifically, Table \ref{tab:overview} examines the key aspects of ingredient substitutions through the lens of four essential questions: who, what, why, and how.

\section{Findings}

\captionof{table}{Analysis of Rule-based Techniques for Ingredient Substitution}
\label{tab:table2}
\begin{longtable}{
  p{0.15\linewidth} p{0.25\linewidth} p{0.25\linewidth} p{0.25\linewidth}
}
\hline
\textbf{Reference} & \textbf{Approach} & \textbf{Dataset and its Size} & \textbf{Goal of Ingredient Substitutions} \\
\hline
\cite{akkoyunlu2017investigating} & Investigating substitutability of food items in consumption data through graph mining approaches & Identifying Substitution pairs the French dataset INCA 21\footnote{\url{https://tinyurl.com/2xbj7mfb}} & To provide personalized dietary recommendations, helping people improve their eating habits \\
\hline
\cite{skjold2017intellimeal} & Reusing domain knowledge in the adaptation process using Case-based reasoning (CBR) & N/A & IntelliMeal: a system customizing recipes to a given user query \\
\hline
\end{longtable}

\captionof{table}{Analysis of Vector Embedding Techniques for Ingredient Substitution}
\label{tab:table3}
\begin{longtable}{
  p{0.15\linewidth} p{0.25\linewidth} p{0.25\linewidth} p{0.25\linewidth}
}
\hline
\textbf{Reference} & \textbf{Approach} & \textbf{Dataset and its Size} & \textbf{Goal of Ingredient Substitutions} \\
\hline
\cite{shidochi2009finding} & Finding replaceable materials in cooking recipe texts in the context of cooking actions through embedding                                           & Recipe website\tablefootnote{\url{http://www.ajinomoto.co.jp/recipe}}\newline Replaceable materials from 391 recipe texts classified into nine groups                                                                                                                     & To propose a method to identify replaceable materials based on characteristic cooking actions derived from a large corpus of recipe text, aiming to meet user demands effectively \\
\hline
\cite{boscarino2014automatic} & Automatic extraction of ingredient's substitutes through topic modeling & Recipe website\footnote{\url{http://www.food.com/recipes##}} \newline 12,515 recipes from a Dutch grocery market website\footnote{\url{http://www.ah.nl/allerhande/recept}}. Dataset containing 2,199 unique ingredients & To alter the recipes based on contextual information, such as culture, sensory perception, and seasons, etc. \\
\hline
\cite{ooi2015ingredient} & Ingredient substitute recommendation for allergy-safe cooking based on food context                                                                   & 100 recipes sourced from the online recipe platform Cookpad\footnote{\url{ https://cookpad.com/us}} & Allergy-safe \\
\hline
\cite{yamanishi2015alternative} & Alternative-ingredient recommendation based on co-occurrence relations & 4508 randomly selected recipes from Recipe database housing a collection of 18,650 recipes \footnote{\url{http://www.kyounoryouri.jp/}} & To provide alternative-ingredients \\
\hline
\cite{achananuparp2016extracting} & Extracting food substitutes from food diary via distributional similarity and co-occurence relations & 2,000 food substitute pairs from CrowdFlower, an online crowdsourcing service CrowdFlower\footnote{\url{http://www.crowdflower.com}}   & Healthier Ingredient recommendation \\
\hline
\cite{shino2016recommendation} & Recommendation system for alternative-ingredients based on co-occurrence relation on recipe database and the ingredient category & Descriptions extracted from recipes sourced from Cookpad, which consists of 1,715,595 Japanese recipes & Ingredients unavailability \\
\hline
\cite{kazama2018neural} & A neural network system for transformation of regional cuisine style through word2vec embedding & The Yummly recipe website dataset consists of 39,774 recipes originating from 20 countries & Local cuisine-based recommendation \\
\hline
\cite{lawo2020veganaizer} & Identifying vegan-friendly substitutions through FastText & Dataset comprising 384,181 recipes, containing a total of 12,196 unique ingredients from two prominent German recipe websites \footnote{\url{https://www.kochbar.de}} \footnote{\url{https://www.chefkoch.de}} & Veganaizer: AI-assisted ingredient substitution for Vegan-friendly substitution \\
\hline
\cite{pan2020food} & Food recipe alternation and generation using recurrent neural networks & Dataset containing 3,433 recipes spanning 15 different cuisine styles gathered from Spoonacular website  \footnote{\url{https://www.spoonacular.com}}   & Local cuisine-based recommendation \\
\hline
\cite{morales2021word} & A word embedding-based method for unsupervised adaptation of cooking recipes & 267,071 recipes scraped from a recipe website such as AllRecipes\footnote{\url{ https://www.allrecipes.com}}, BBC Food Recipe\footnote{\url{https://www.bbc.co.uk/food/recipes}}, CookStr\footnote{\url{https://www.cookstr.com}}, Epicurious\footnote{\url{https://www.epicurious.com}} & Vegan-friendly substitution \\
\hline
\cite{pacifico2021ingredient} & Ingredient substitute recommendation using collaborative filtering & 230,000 recipes and over 1 million user interactions such as comments and ratings, collected from 2000 to 2018 from the Food.com. Dataset comprises 24,105 recipes, featuring 8,633 unique ingredients across both categories & Allergy-safe recipe generation  \\
\hline
\cite{pellegrini2021exploiting} & Clustering food embeddings for ingredient substitution and Image to recipe generation & Identifying Substitution pairs from open-source dataset Recipe1M+ (\cite{marin2021recipe1m+}) & To provide substitute recommendations in dietary use cases \\
\hline
\cite{pacifico2022automatic} & Recipe ingredient substitution based on text mining and data clustering approaches & Dataset was extracted from the Food.com website & Local cuisine-based recommendation \\
\hline
\cite{pacifico2022improved} & Average support value for ingredient substitute recommendation through collaborative filtering & N/A & Ingredients unavailability \\
\hline
\cite{antognini2022assistive} & Assistive recipe editing through critiquing using recurrent neural networks, generative language models and auto-encoders & Identifying 1,488 substitution pairs from open-source dataset Recipe1M (\cite{salvador2019inverse}) & To create recipes that satisfy user constraints \\
\hline
\cite{fatemi2023learning} & Clustering based on food embeddings and  image to recipe generation & Recipe1Msubs containing 6,653 ingredients based on Recipe1M & Allergy-safe \\
\hline
\cite{loesch2024automated} & Leveraging graph neural networks and Nutri-scores for ingredient substitution & -Build the knowledge graph using diverse information on food items, nutrients, products, flavor molecules, tags, and substitutions from USDA, OpenFoodFacts, FooDB, Food Tags, and Ground Truth \newline
-Ground Truth dataset (\cite{loesch2022automated}) with 370 unique food items and 704 substitutions
 & To recommend healthier food ingredient substitutions with better nutritional profiles \\    
\hline
\end{longtable}

\captionof{table}{Analysis of Knowledge graph Techniques for Ingredient Substitution}
\label{tab:table4}
\begin{longtable}{
  p{0.15\linewidth} p{0.25\linewidth} p{0.25\linewidth} p{0.25\linewidth}
}
\hline
\textbf{Reference} & \textbf{Approach} & \textbf{Dataset and its Size} & \textbf{Goal of Ingredient Substitutions} \\
\hline
\cite{shirai2020semantics} & Semantics-driven ingredient substitution in the FoodKG using graph traversal methods & -Their own dataset (a set of 2,300 substitutions pairs from The Cook's Thesaurus (i.e., pairs of “target ingredient” to “substitute ingredient”) \footnote{\url{ foodsubs.com}} \newline -1,161 unique ingredients providing substitutions for 928 target ingredients
 & Allergy-safe \\
\hline
\cite{shirai2021identifying} & Identifying ingredient substitutions using a knowledge graph of food using custom heuristics & AllRecipes, Food Network, Colorado State University (CSU), and North Dakota State University (NDSU). 137(Substitution pairs), The Cook's Thesaurus 2,360 (Substitution pairs), 1,004 (Unique ingredients), 3,313 (Substitution pairs), 1,331 (Unique ingredients) (\cite{majumder2019generating}) & Healthier ingredient recommendation: Low-carb recommendation \\
\hline
\cite{li2022food} & Learning knowledge graph embeddings with adversarial substitutions & FoodKG (\cite{haussmann2019foodkg}) & A collection of adversarial sample generation strategies for different food substitutions \\
\hline
\cite{loesch2022automated} & Ingredient substitution using knowledge graph embeddings & 3344 substitution pairs sourced from Food.com reviews, linked to USDA & Healthy ingredient alternatives with similar nutritional characteristics \\
\hline
\end{longtable}

\captionof{table}{Analysis of Theoretical Approach for Ingredient Substitution Design for Ingredient Substitution}
\label{tab:table5}
\begin{longtable}{
  p{0.15\linewidth} p{0.25\linewidth} p{0.25\linewidth} p{0.25\linewidth}
}
\hline
\textbf{Reference} & \textbf{Approach} & \textbf{Dataset and its Size} & \textbf{Goal of Ingredient Substitutions} \\
\hline
\cite{saengsupawat2014ontology} & Theoretical rules for knowledge acquisition for Thai ingredient substitution & N/A & Ingredients unavailability \\
\hline
\cite{gaillard2015improving} &Ingredient substitution using formal concept analysis and adaptation of ingredient quantities with mixed linear optimization & N/A & Ingredients unavailability \\        
\hline
\cite{gaillard2017adaptation} & Adaptation of TAAABLE to the CCC'2017 through formal concept analysis & N/A & Managing available ingredients for salads and cocktails \\
\hline
\cite{lawrynowicz2022food} & Ontological design patterns for ingredient substitution &  N/A & Design considerations while developing ontology for ingredient substitution \\
\hline
\cite{bikakis2023graphical} & Formal reasoning approaches for recipes & N/A & To propose formal definitions for recipe substitutions: missing ingredients, unfeasible actions (e.g., lack of equipment, time, or skill), or product changes\\
\hline
\end{longtable}

\subsection{Dataset for Ingredient Substitution - RQ1}
Unlike benchmarks that are available for recipe datasets, it is uncommon to find benchmark datasets for ingredient substitution. Several works create ingredient substitution pairs from recipe datasets based on co-occurence relations. Additionally, there are various websites, such as foodsubs.com \footnote{\url{https://foodsubs.com/}}, that provide ingredient substitution information independent of specific recipes. They may include nutrition information and contextual information such as vegetable substitution, dairy substitution and so on. A few works rely on the comments from recipe websites where users generally suggest ingredient substitution, which can be considered as gold standard as it is provided by domain experts. To add, these ingredient substitutions require capturing the recipe context which is challenging to curate and is a rarity. We also investigate these data sources in terms of their trustworthiness in section 5.2. Given diet is a high stake domain, it is essential that these substitutions and dietary labels are from trusted sources. In this section, we present the findings on possible sources and approaches utilized for ingredient substitution dataset curation.

\subsubsection{Recipe websites}
Recipe websites serve as crucial resources for researchers in the field of food computing, particularly for those focusing on ingredient substitution techniques. For instance, \cite{shidochi2009finding} utilized a dataset of cooking recipe texts, which were input as HTML files from a specific recipe site providing Japanese food recipes, to find replaceable materials in recipes. Similarly, \cite{boscarino2014automatic} leveraged a dataset from a Dutch grocery market website comprising 12,515 recipes to explore suitable contexts for substitution rules. The dataset was pruned to enhance the relevance and efficiency of the analysis, resulting in a reduced ingredient space containing 2,199 unique ingredients. \cite{ooi2015ingredient} gathered a dataset of 100 recipes from the online recipe platform Cookpad, specifically selecting recipes that included common allergy-inducing ingredients. These datasets from recipe websites were instrumental in developing systems that can recommend ingredient substitutions based on various factors such as allergy safety, cultural context, and seasonal availability.

\subsubsection{Comments from recipe websites}
Comments from recipe websites can provide valuable insights into user preferences, substitutions, and cooking experiences. These user-generated contents are rich sources of data for researchers aiming to understand real-world cooking practices and develop systems that can adapt recipes to meet specific dietary needs or ingredient availability. For example, \cite{pacifico2021ingredient} utilized over 1 million user interactions, such as comments and ratings, from the Food.com website to recommend alternative recipes for recipes containing forbidden ingredients. This approach allows for the adaptation of recipes by replacing ingredients with safe alternatives, taking into account user food restrictions. The analysis of comments and user feedback is essential for creating personalized and context-aware recommendation systems.

{\subsubsection{Manually curated datasets for Ingredient Substitution Guidelines}
In the field of ingredient substitution, complete datasets are scarce compared to those available for recipes. However, there is a growing body of datasets derived from substitution guidelines and expert recommendations. For instance, \cite{shirai2021identifying} meticulously assembled a dataset from various reputable sources such as AllRecipes, Food Network, Colorado State University (CSU), and North Dakota State University (NDSU). These sources offer valuable insights and expert advice on substituting ingredients in recipes, forming the basis of curated datasets for further research and analysis.

\subsubsection{Identifying Substitution pairs from open-source datasets}
Identifying substitution pairs from existing datasets is one of the common approaches to create datasets for ingredient substitution recommendation systems. For instance, \cite{loesch2024automated} utilized datasets from various sources, including the USDA and OpenFoodFacts, to develop an automated system to recommend healthier food substitutions[1]. By analyzing substitution pairs and the nutritional profiles of ingredients, these systems can provide users with healthier alternatives that align with their dietary goals. In the meanwhile, \cite{shirai2021identifying} extracted substitutions from 1.1 million user reviews of Food.com recipes published by \cite{majumder2019generating}. Then, the substitutions were parsed from the user reviews using simple patterns indicating that a substitution was made in the recipe, such as “substitute a for b” or “replace b with a”. 

\subsubsection{Summary}
The datasets highlighted in the provided tables \ref{tab:table2}, \ref{tab:table3}, and \ref{tab:table4} are diverse and serve as the foundation for various research studies in the field of ingredient substitution. Primarily sourced from recipe websites and user-generated content, such as comments and ratings, these datasets provide a rich context for understanding ingredients with respect to cooking practices and food preferences. They vary significantly in size and scope, ranging from hundreds to millions of recipes. Additionally, they include both general and specialized collections, such as those focusing on allergy-safe cooking or regional cuisine styles. The diversity of these datasets reflects the complexity of the culinary domain and the varied approaches researchers take to address the challenge of ingredient substitution.

\subsection{Computational Techniques for Ingredient Substitution - RQ2}
\subsubsection{Rule-based systems}Rule-based systems refer to systems that make decisions or perform actions based on a predefined set of rules for ingredient substitution. These rules are typically expressed in the form of a flowchart incorporating 'if-then' statements. Rule-based systems work by evaluating input data against a set of rules and then executing the actions associated with the rules that match the input conditions. In the examples provided below, both graph mining approaches and case-based reasoning systems utilize rules to make decisions. One of the key advantages of rule-based systems is their transparency and explainability. Since each decision is based on explicit rules, it is easier for users to understand why a certain recommendation was made. They also provide a structured framework for decision-making, which can be advantageous in domains where consistency and clarity are important. However, rule-based systems may not be as scalable or generalizable compared to other approaches. Managing a large number of rules can become complex and cumbersome, especially as the system grows in size and complexity. Table \ref{tab:table2} provides an overview of the studies that follows rule-based approaches.
\begin{itemize}
\item
\textit{Graph Mining Approaches}: Through graph mining techniques, their model retrieves sets of dietary contexts and substitutable sets. In graph mining, a clique is defined as a node that is completely connected, creating a maximal clique, and the Bron-Kerbosch algorithm is employed to search for these maximal cliques. Based on this substitution rule to find cliques, \cite{akkoyunlu2017investigating}'s research computes substitutional scores.

\item
\textit{Case-Based Reasoning}: In a similar vein, \cite{skjold2017intellimeal}'s system known as IntelliMeal consists of twenty-one sandwich recipe cases, along with hierarchical taxonomies and a set of adaptation rules, including the simple rules and substitution rules. They utilize the domain knowledge stored within a Case-Based Reasoning (CBR) system for adapting recipes in response to user queries. For the retrieval process, the system compares each user query to all the cases in the case base. It examines each case individually, going through each attribute in the query, while disregarding any undefined attributes.
\end{itemize}

\subsubsection{Vector Embedding-based systems}Vector embedding-based systems utilize numerical vector representations that encapsulate the connections and meaning of words, phrases, and various data entities. These systems operate by encoding this information into these numeric vectors, allowing for mathematical operations such as similarity calculation or pattern recognition. For instance, in the context of recipe analysis, researchers utilize vector embeddings to quantify relationships between ingredients. Vector embedding-based systems excel at pattern mining, enabling tasks like ingredient substitution and recipe generation based on learned associations. However, their reliance on complex mathematical representations makes them challenging to interpret and understand. Despite this lack of transparency, their effectiveness lies in their ability to generalize well across diverse datasets due to their robust pattern mining capabilities. Table \ref{tab:table3} provides an overview of the studies that follows vector embedding approaches.

\begin{itemize}
\item
\textit{Topic Modeling}: \cite{boscarino2014automatic} proposed a statistical model used for topic modeling using Latent Dirichlet Allocation (LDA) to rank ingredients based on their probability for replaceability in each recipe. 

\item 
\textit{Co-occurence Relations}: Many researchers employ vector embedding-based approaches, and among them, co-occurrence relations within a recipe database are commonly used (\cite{yamanishi2015alternative, achananuparp2016extracting}). Co-occurrence relations quantify how frequently certain combinations of ingredients appear together in recipes. \cite{achananuparp2016extracting}'s research is grounded in the distributional hypothesis from linguistics, which assumes that words appearing in similar contexts tend to convey similar meanings. For ingredient substitution, it is assumed that foods consumed in similar contexts can serve as substitutes for each other.

\item
\textit{FastText}: Using the embedding algorithm FastText, their model ranks a list of substitutes. Similar to Word2Vec, FastText employs a neural network to predict a target word based on its surrounding context. However, FastText decomposes words into character n-grams (subwords) and represents each word as a combination of these embedded subword vectors. FastText effectively handles out-of-vocabulary (OOV) words and infrequent terms, as it can construct embeddings based on recognized subword components. This capability expands the applicability of word embeddings beyond the limitations of vocabulary size observed in Word2Vec. 

\item 
\textit{Recurrent Neural Networks}: Alongside vector embeddings, many researchers leverage language models to generate revised recipes incorporating ingredient substitution (\cite{pan2020food, antognini2022assistive}). In order to address the unavailability of essential recipe ingredients, \cite{pan2020food} employed Skip-gram models to represent ingredients and recipes as vectors and measured the degree of similarity among words through cosine similarity. From long sequences of stored text, they utilized language models including N-gram and Recurrent Neural Networks (RNN) with Long Short-Term Memory (LSTM) layers as a neural network model for recipe generation. 

\item
\textit{Generative Models}: \cite{antognini2022assistive} introduce RecipeCrit, a denoising-based model designed to complete recipes and gain a deeper understanding of the semantic relationships between ingredients and instructions. A novelty in this work is the introduction of an unsupervised critiquing approach, which enables users to provide ingredient-centric feedback iteratively.

\item
\textit{Clustering}: While most research heavily relies on text data, some researchers pay attention to the prowess of a multimodal approach, such as integrating text data with images (\cite{pellegrini2021exploiting, fatemi2023learning}). \cite{pellegrini2021exploiting} present Food2Vec and FoodBERT for ingredient embeddings, which are combined with images, and they determine which ingredients are substitutable by using a clustering algorithm. While \cite{pacifico2022automatic}'s research also employs data clustering, their research solely relies on text data to address ingredient unavailability in creating customized recipes. While the novelty of \cite{pellegrini2021exploiting}'s research lies in the multimodal representation of inputs and outputs, like other vector embedding approaches, their approach may face feasibility challenges due to its context-free nature. For instance, if they disregard essential contexts such as cooking actions, it becomes challenging to generate useful recipes and can make it difficult for individuals to adapt their recipes to their dietary needs.

\item
\textit{Collaborative Filtering}: Another line of study for ingredient substitution is collaborative filtering (\cite{pacifico2021ingredient, pacifico2022improved}). Basically, collaborative filtering is a recommendation system that includes or excludes certain items for recommendation based on users' preferences and behaviors of similar users. \cite{pacifico2021ingredient} depart from traditional collaborative filtering and employ an advanced approach called Average Support Value (ASV) filters, which is a type of collaborative filtering technique used to fine-tune the quality of recommendations by leveraging information from users' average support values.

\item
\textit{Image to recipe generation}: \cite{pellegrini2021exploiting} introduce a cutting-edge language model that integrates image-based embeddings, thereby enhancing the model's robustness through multimodal capabilities. Similarly, \cite{fatemi2023learning} present a state-of-the-art personalized image-to-recipe generation system leveraging vector embeddings. Their approach employs the Graph-based Ingredient Substitution Module (GISMo) to recommend alternative ingredients for revised recipes, marking a significant advancement in personalized recipe generation. Specifically, \cite{fatemi2023learning} introduce a novel personalized inverse cooking pipeline wherein GISMo utilizes predicted ingredients alongside user-specified substitutions to craft personalized recipes from images. This research stands at the forefront of the field by incorporating multimodal inputs and leveraging vector embeddings to enhance contextual understanding and relational information for more personalized recipe recommendations.
\end{itemize}

\subsubsection{Knowledge Graph-based systems}
Recently, the utilization of knowledge graphs has emerged as promising approaches to address the intricate challenges associated with ingredient substitution within food computing domains. Knowledge graphs, such as FoodKG, encapsulate abundant semantic relationships among various culinary entities, thereby facilitating comprehensive analyses and enabling informed decision-making processes for ingredient substitutions. To be specific, they can capture relationships among several real-world entities in various semantic contexts. In addition, in the context of AI, knowledge graphs facilitate human-like reasoning, including inference and explainability. Through inferencing using knowledge graphs new information can be derived from existing facts and relationships. This section examines methodological approaches within knowledge graph-based systems, accentuating their contributions and advancements in this field. Table \ref{tab:table4} provides an overview of the studies that follow knowledge graph based approaches.

\begin{itemize}
\item
\textit{Graph Traversal Methods}: \cite{shirai2020semantics} are the first to leverage the capabilities of FoodKG, a knowledge graph containing recipe and ingredient data established by \cite{haussmann2019foodkg}. Through this knowledge graph, \cite{shirai2020semantics} aim to identify substitutable ingredients while considering nutritional information. Previous research solely based on vector similarity has overlooked nutritional information. However, \cite{shirai2020semantics} propose a system that ranks potential substitutable ingredients using the knowledge graph. They utilize forms of word embeddings, such as Word2Vec Similarity, SpaCy NLP Similarity (\cite{spacy2}), Recipe Context Similarity, and Ingredient Pairing Similarity, as well as the linked ontology of food from FoodOn. For example, consider the “Mashed Potato Recipe,” which includes three ingredients: “Unsalted butter,” “Potato,” and “Salt.” For the linked data related to “Unsalted Butter,” it is matched with USDA's nutritional information (“Butter, Unsalted”: Calories: 717, Sodium: 11 mg, Total Fat: 81 g) and with FoodOn's ontology (butter(unsalted) -> cow milk butter food product (subclass) -> cow milk based food product (subclass)). For each ingredient of the target recipe, the system traverses its linked FoodOn class in the FoodKG. If a path exists between the prohibited food class and the ingredients, the system rules out those ingredients for substitutions.

\item 
\textit{Custom Heuristics}: As a follow-up research, in “Identifying ingredient substitutions using a knowledge graph of food,” \cite{shirai2021identifying} proposed an advanced approach. To extract the implicit semantic information unique to FoodKG's dataset, they employ the Word2Vec model trained on ingredient names and recipe instructions from Recipe 1M. Additionally, \cite{shirai2021identifying} utilize SpaCy's word embedding model. Essentially, they calculate the cosine similarity between ingredient names. The cosine similarity between the Word2Vec and SpaCy embeddings of ingredients a and b is denoted as Wa,b and Sa,b, respectively. Furthermore, they derive a vector indicating the likelihood of ingredient co-occurrences within a recipe alongside ingredient j. This vector is then used to compute the co-occurrence substitutability score Da,b. \cite{shirai2021identifying} also employ Pa,b for computing the positive pointwise mutual information (PPMI) for each ingredient and recipe context. For example, if a recipe includes bread, peanut butter, and jelly, it can be inferred that jelly is in the context of (bread and peanut butter). By amalgamating these four numerical metrics, they create a formula named Diet-Improvement Ingredient Substitutability Heuristic (DIISH) to rank ingredients. However, they do not elucidate the criteria or rationale behind combining these metrics.

\item
\textit{Knowledge Graph Embeddings}: More recent research employs Knowledge graph embeddings (KGEs) (\cite{li2022food, loesch2022automated}). KGEs transform each entity and relation within a knowledge graph into vector dimensions, known as embedding dimensions. These embeddings facilitate the manipulation of graph elements, such as entities and relations, enabling effective prediction tasks like entity classification and link prediction. By embedding knowledge graph elements into a continuous vector space, KGEs enhance the efficiency and accuracy of various knowledge-driven tasks within the graph. \cite{li2022food}'s approach underscores the importance of contextual information within food KGs by employing a pre-trained language model to represent entities and relations. Their model is trained for two critical tasks: predicting a masked entity within a given KG triple and evaluating the plausibility of such triples. This research extends beyond previous studies on ingredient substitution, which often focus on semantic similarity while neglecting the essential element of context. As another research to use KGEs, the main purpose of \cite{loesch2022automated}'s research is to assist individuals in making ingredient substitutions that have similar nutritional values to the target replaced ingredients. They employ KGEs to develop a food ingredient recommendation for healthier ingredient substitutions. Their research builds upon the research of \cite{shirai2021identifying}, which proposes a model providing a ranking system for ingredient substitutions based on a knowledge graph that includes semantically interlinked web data. \cite{loesch2022automated}'s approach aims to provide a broader range of ingredient recommendations by utilizing knowledge graph embeddings. To recommend ingredients with similar nutritional values to the target replaced ingredients, they construct a knowledge graph using the OpenFoodFacts dataset, which is a food product database, and the USDA, which offers nutritional information for food products. Their knowledge graph embedding model is designed to capture semantic similarities. For example, the more connected ingredients are by relations, the closer they are in terms of nutritional information. They establish relationships between food knowledge by linking OpenFoodFacts and USDA. Additionally, based on the U.S. FDA's Recommended Dietary Allowances (RDAs), they tag the ingredients on the knowledge graph.
\end{itemize}

\subsubsection{Theoretical Approaches for Ingredient Substitution Design}
There exists another line of research to offer the ingredient substitution designs in terms of ontology designs (\cite{saengsupawat2014ontology, gaillard2015improving, gaillard2017adaptation, lawrynowicz2022food, bikakis2023graphical}). On the one hand, food ontologies represent a structured knowledge base encompassing various food-related concepts. They categorize ingredients based on diverse properties, including nutritional content (proteins, carbohydrates, fats), sensory attributes (taste, texture), and origin (plant-based, animal-based, processed). The ontological structure complements semantic similarity analysis by adding layers of constraint. For example, while beef and vegetarian patties might possess semantic similarity, their distinct class labels within the ontology (e.g., “animal-based” vs. “plant-based protein”) prevent their erroneous recommendation as direct substitutes. However, designs proposed for food ontologies also have limitations. They may not adequately capture the nuances of different cooking methods, which can significantly impact ingredient functionality and suitability. The textural and functional properties of an ingredient might change based on cooking techniques like deep-frying, stir-frying, or boiling, potentially necessitating a more context-specific substitution recommendation. Table \ref{tab:table5} provides an overview of studies that discuss design considerations.

\begin{itemize}
\item
\textit{Rule languages}: The main purpose of \cite{saengsupawat2014ontology}'s research is to find substitutes for rare ingredients that are not available in certain locations while preserving their original sensory properties. Their proposed model employs domain ontology to design classes and properties for each class, such as taste, flavor, and textures. Then, the model applies the Semantic Web Rule Language (SWRL) to incorporate rule bases into the domain ontology for ingredient substitution inference.

\item 
\textit{Formal Concept Analysis}: In the context of the 2015 Computer Cooking Contest (CCC) expanded the TAAABLE system in two significant ways (\cite{gaillard2015improving, gaillard2017adaptation}). Firstly, they employed Formal Concept Analysis (FCA), a mathematical framework for analyzing complex data, to tackle the challenge of unavailable ingredients. FCA, which emphasizes conceptual clustering, knowledge discovery, and data relationships, finds extensive application in knowledge representation, data analysis, and information retrieval. Secondly, \cite{gaillard2015improving} refined ingredient quantities to better align with real-life cooking scenarios. For instance, they expressed the quantity of ingredients like lemon in easily understandable terms for humans, such as ‘a quarter’ or ‘a half,’ rather than less intuitive measurements like ‘54 g,’ which correspond to half a lemon. This enhancement in quantity adjustment was accomplished through the application of mixed linear optimization techniques.

\item
\textit{Ontological Design Patterns}: \cite{lawrynowicz2022food} propose a systematic ontology design pattern for ingredient substitution. They explain the concept of ingredient substitution in food recipes and present the following specific research questions: “What related concepts should be taken into account to define the substitution's context (e.g., conditions, goals)? How to link the proposed model to existing food models and recommended design patterns?” (\cite{lawrynowicz2022food}). This paper does not have a specific goal of food ingredient substitution, such as catering to individuals who require dietary modifications due to health conditions like diabetes (\cite{seneviratne2021personal}). \cite{lawrynowicz2022food}'s paper argues that their ontology design pattern can serve as a prerequisite for constructing AI-based models for ingredient substitution, as well as for integrating other concepts into the model, such as food ontology and knowledge graph. A good aspect of this article is its capability to provide a comprehensive understanding of substituting food ingredients. For instance, the article sheds light on the potential necessity to adjust the recipe when ingredients are altered. Categorized under three headings—tastiness, dietary needs, and technical changes in the cooking process—they present an ontological design pattern for ingredient substitution. Furthermore, they assess the alignment of the developed pattern with existing food ontologies based on the FoodOn ontology. 

\item
\textit{Formal Reasoning}: To graphically formalize recipes for use within a reasoning framework, we propose a set of formal definitions that not only simplify but also provide clarity to various aspects of recipe analysis (\cite{bikakis2023graphical}). These definitions encompass the comparison of recipes, the composition of recipes their fundamental building blocks, and the deconstruction of complex recipes into their constituent subrecipes. Additionally, \cite{bikakis2023graphical}'s research introduces and evaluates two distinct formal definitions for recipe substitution, a vital component when dealing with missing ingredients, unfeasible actions, or the need to modify the final culinary outcome. Specifically, in terms of ingredient substitution, their research employs formal techniques for substituting ingredients, actions, and even subrecipes within recipes. This comprehensive approach underscores the strength of their paper and its contribution to the field of formalism in ingredient substitution through recipe revision. It addresses challenges such as ingredient unavailability, dietary restrictions, environmental considerations, or the inability to execute specific actions. Notably, the relevance of this work extends beyond formalism, as it finds common ground with research in food ingredient substitution, bridging the gap between formal reasoning and practical culinary applications. 
\end{itemize}

\subsubsection{Summary}
Neuro-symbolic approach, which combines deep learning with knowledge graphs, offers a powerful approach to recommending food ingredient substitutions. Each method complements the other. Deep neural networks excel at identifying ingredient substitutes by recognizing complex patterns in large datasets. However, deep learning models often act as "black boxes," making their decisions difficult to interpret or explain. Similarly, knowledge graphs provide a human-readable, symbolic approach, using semantic rule-based systems to recommend ingredient substitutions, but they lack ability to generalize. By integrating deep learning's pattern recognition capabilities with the explainable, rule-based reasoning of knowledge graphs, this combined approach could enhance the accuracy of ingredient substitution recommendations and provide a better understanding of why each recommendation is made. For instance, a deep learning model could identify potential substitutes based on recipe contexts and ingredient co-occurrences, while a knowledge graph could refine these suggestions by applying nutritional constraints and culinary rules. In conclusion, while each method category has its strengths and limitations, the future of ingredient substitution systems likely lies in hybrid approaches that leverage the complementary strengths of multiple techniques. By combining the pattern recognition capabilities of vector embeddings, the semantic reasoning of knowledge graphs, and the explainability of rule-based systems, researchers can develop more robust and versatile ingredient substitution models that cater to diverse user needs and preferences.

\subsection{Additional Knowledge for Ingredient Substitution - RQ3}
This section presents the findings on the additional knowledge of ingredients that are considered by existing research studies while modeling a system to identify a suitable substitution. To provide ingredient substitution recommendations based on the user’s health condition and food preferences, it is necessary to include health labels of an ingredient, such as vegan or lactose-free, and disease contexts such as the glycemic index of the ingredient for diabetes patients. Additionally, AI models can learn implicit patterns from visual representations of ingredients. For example, \cite{akkoyunlu2017investigating} aim to identify food pair substitutions based on contextual information rather than nutritional information. They define two contexts: the dietary context and the food intake context. The dietary context of a food item is determined by the other food items consumed with it. For example, in the meal {coke, burger, fries}, the dietary context of the burger is {coke, fries}. The food intake context encompasses variables such as meal types, locations {home, workplace, restaurant}, participants {family, friends, coworkers, alone}, and other relevant factors.

\subsubsection{Nutrition of Ingredient} Some research (\cite{shirai2020semantics}) considers the nutritional content of each ingredient, especially when using their nutrition information. To accommodate various dietary needs, many systems compute the nutritional values of all ingredients within the target recipes. Subsequently, it verifies whether these calculated values adhere to any dietary restrictions pertinent to individuals' health objectives.

\subsubsection{Health Labels} 
Health labels, such as vegan, lactose-free, or gluten-free, are diverse. For instance, \cite{shirai2021identifying} also include health labels like high or low carb, which are derived from nutritional data such as that provided by the U.S. Department of Agriculture (USDA). These labels provide valuable information to users seeking healthier or more suitable ingredient options, facilitating more personalized and informed food choices.

\subsubsection{Recipe Context} Recipe context refers to identifying ingredient substitutions within the context of a recipe. For example, the substitute of an egg in a cake could be yogurt or vegetable oil. On the other hand, the substitute of egg in a pastry like croissant could be milk where egg is used to glaze the croissants before baking. Ingredient substitution models can utilize recipe contexts to enhance the accuracy of recommending substitutes. For example, \cite{shirai2020semantics, shirai2021identifying} employ recipe context similarity metrics to provide more accurate and relevant ingredient substitutions for each ingredient and its context within a recipe. Additionally, to provide more context-relevant substitutions, \cite{pacifico2021ingredient} propose using collaborative filtering combined with recipe context. Their model offers allergy-safe ingredient substitution recommendations to facilitate the automatic generation of allergy-safe recipes.

\subsubsection{Visual Context} Visual context refers to the use of ingredient images or recipe images to recommend suitable ingredient substitutions. \cite{pellegrini2021exploiting} employed language models for ingredient substitutions by integrating ingredient image-based embeddings. Meanwhile, \cite{fatemi2023learning} proposed personalized image-to-recipe generation models using vector embeddings. Both proposed models show that incorporating ingredient or recipe images combined with vector embedding techniques enhances accuracy for ingredient substitution.

\subsubsection{User’s health and food preferences} \cite{loesch2022automated} point out the limited range of dietary constraint categories provided by \cite{shirai2021identifying}, such as vegetarian or allergens for peanuts, as well as the limited consideration of nutritional information, which \cite{shirai2021identifying} only consider as high or low carb nutritional information.

\subsubsection{Flavor of Ingredients} Flavor is a crucial feature to consider in ingredient substitution because people do not eat food solely for its nutritional value; flavor enhances their enjoyment of eating. \cite{kazama2018neural} incorporate flavor information in their proposed model for recommending local cuisine. They consider combinations of flavor components to capture and preserve the characteristics of specific localized foods while recommending ingredient substitutions.

\subsubsection{Summary} The existing works in the field of food ingredient substitution have not comprehensively leveraged the available contextual information related to diseases, health labels, and flavor profiles of ingredients. This could be due to the fact that most of the existing work evaluate ingredient substitution on “healthier” substitutions or based on ingredient replaceability. Furthermore, while some works have developed vegan systems (\cite{lawo2020veganaizer}), other health labels associated with ingredients have not been utilized. Incorporating health labels could enable the generation of recipes tailored to specific dietary requirements or preferences, such as vegan, gluten-free, or low-sodium diets such as \cite{seneviratne2021personal}. Additionally, the flavor profiles of ingredients, available from resources like FlavorDB, have not been considered by all existing works. Considering flavor profiles could lead to the generation of recipes with more balanced and appealing flavor combinations, enhancing the overall culinary experience. Addressing these gaps by incorporating such contextual information could potentially improve the relevance, suitability, and appeal of the generated recipes, catering to diverse dietary needs and preferences.

\subsection{Ingredient substitution: Applications - RQ4}
The need for ingredient substitution might include several cases such as missing ingredients, dietary restrictions, or dietary constraints. \textit{Dietary restrictions} are conditions necessitated by health-related issues, such as lactose intolerance, allergens, or celiac disease that requires a gluten-free diet. An incorrect recommendation in this case could be detrimental. \textit{Dietary constraints} encompass personal dietary choices influenced by cultural, religious, or lifestyle factors, such as veganism, vegetarianism, or ketogenic diets. While most research aims to provide ingredient recommendations to comprehensively satisfy users' preferences and dietary constraints, some research narrows down its scope to address specific applications. There are five applications to consider in food ingredient substitution research, including vegan-friendly substitution, allergy-safe options, healthier food recommendations, ingredient unavailability, and local cuisine-based substitutions. It is to be noted that the sources of knowledge for these constraints are diverse, including nutritional databases. For example, determining whether an ingredient is vegan may involve referencing databases or sources that catalog ingredients derived from animal products. By incorporating these dietary constraints into the recommendation process, researchers aim to provide users with personalized and relevant ingredient substitutions that align with their dietary preferences and restrictions.

\subsubsection{Vegan-friendly substitution} In the pursuit of vegan-friendly ingredient substitutions, \cite{lawo2020veganaizer} introduced the veganizer, a system dedicated to transforming recipes into plant-based and vegan-friendly alternatives through ingredient substitutions. Likewise, \cite{morales2021word} proposed a model that takes into account nutritional preferences, adapts to comparable ingredients, and accommodates vegetarian and vegan dietary restrictions.

\subsubsection{Allergy-safe} Due to the increase in food-allergic patients, allergy-safe ingredient substitutions and recipes are in need. However, the number of allergy-safe recipes available is less than that of normal recipes. To address this issue, \cite{ooi2015ingredient} proposed an approach to recommend allergy-free ingredient substitutions extracted from a large recipe database. \cite{shirai2020semantics} provided ingredient substitutions addressing personal preferences, allergies, and nutritional or other dietary considerations. With the similar purpose, using collaborative filtering and recipe context, \cite{pacifico2021ingredient}'s model proposed allergy-safe ingredient substitution recommendations to generate automatic allergy-safe recipe generation. In a similar vein, \cite{fatemi2023learning} proposed recipe personalization through ingredient substitution to cover users' dietary constraints and preferences, such as allergens.

\subsubsection{Healthier Ingredient recommendation} Some research employs the phrase “healthier” food ingredient substitution and provides models to achieve their goals (\cite{achananuparp2016extracting, shirai2021identifying, loesch2022automated}). To be specific, \cite{achananuparp2016extracting} recommend similar and healthier ingredient substitutions based on users' current dietary needs and preferences. \cite{shirai2021identifying}'s approach determines which ingredients are healthier than others based on nutritional information and food classification constraints. Similarly, \cite{loesch2022automated} find healthier ingredient substitutions among ingredients with similar nutritional characteristics. However, the definition of a 'healthier' ingredient in the current research appears to be subjective.

\subsubsection{Ingredients unavailability} One challenge in cooking arises when certain ingredients are not available. To address this issue of ingredient unavailability, some researchers have proposed ingredient recommendation solutions (\cite{saengsupawat2014ontology, gaillard2015improving, shino2016recommendation}).
For example, \cite{saengsupawat2014ontology} proposed a system to deal with the unavailability of rare ingredients by employing a domain ontology combined with a rule-based approach. Similarly, \cite{gaillard2015improving} overcome unavailability problems using formal concept analysis. Meanwhile, \cite{shino2016recommendation} try to solve the problem with a more traditional approach, such as co-occurrence relations in a recipe database and ingredient categories.

\subsubsection{Local cuisine-based recommendation} Researchers have explored various methods to enhance culinary creativity and flexibility in recipe adaptation. \cite{kazama2018neural} introduced a technique to transform a recipe into any selected regional style, whether it be Japanese, Mediterranean, or Italian, aiming to imbue new recipes with the authentic flavors of specific cuisine styles. Building upon this concept, \cite{pan2020food} proposed a method to generate entirely new recipes that capture the essence of a chosen culinary tradition. Meanwhile, \cite{pacifico2022automatic} focused on adapting given recipes by seamlessly replacing unavailable ingredients, ensuring that cooks can still create delicious dishes even when faced with ingredient constraints. These approaches collectively contribute to expanding the repertoire of culinary possibilities and empowering chefs to explore diverse flavors and cuisines with confidence.

\subsubsection{Low-carb recommendation} In their study, \cite{shirai2021identifying} present a use case for a low-carb recommendation system that utilizes a substitute filtering and ranking approach. This system aims to offer suitable substitutes for users' queries, particularly focusing on low-carb options. For instance, when a user searches for a "potato substitute" with an emphasis on low-carb options, their model suggests ingredients that have fewer carbohydrates by ranking them. They employ recipe data in FoodKG to incorporate the USDA’s nutritional information to estimate carbohydrate counts. Then, by searching for low-carb potato substitutes on the web, they provide desirable alternatives by ranking them based on their carbohydrate content.

\subsubsection{Summary} A significant insight is the diverse range of applications and motivations driving research in the food computing field. This survey covers various aspects of ingredient substitution, including vegan-friendly alternatives, allergy-safe options, healthier food recommendations, solutions for ingredient unavailability, and local cuisine-based substitutions. This diversity reflects the complex nature of dietary needs and preferences in modern society, ranging from health-related restrictions to personal lifestyle choices.
A potential gap in current research lies in the subjective nature of certain substitution criteria, particularly in the realm of "healthier" ingredient recommendations. While several studies propose models for healthier substitutions, the definition of what constitutes a "healthier" ingredient appears to lack standardization across different research efforts. This suggests an opportunity for future research to establish more objective and comprehensive criteria for determining the healthiness of ingredient substitutions, perhaps by incorporating a wider range of nutritional factors beyond just calorie or macronutrient. Additionally, there may be room for more interdisciplinary approaches that combine nutritional science, culinary expertise, and data-driven methodologies to create more robust and personalized substitution recommendations.

\subsection{Safety of Ingredient Substitution Systems - RQ5}
While there are several metrics to evaluate a system for safety (\cite{sanneman2022situation, morales2023toward}), in this survey we focus on two major aspects namely trustworthiness and transparency. Not all the studies reviewed in this survey paper cover both trustworthiness and transparency. These two aspects are crucial for evaluating the safety and reliability of ingredient substitution models, yet they are not universally addressed across all studies. Given a high stake and high risk domain like diet, it is essential that trusted information sources are used to train the models or to perform inference on ingredient substitution pairs. In this work, we aim to address AI safety in three aspects, namely, ability of the model to explain the outcome, adherence to medical guidelines and trustworthiness of the data sources and knowledge sources utilized to train the model. 

\subsubsection{Explainablity of the Models} 
Presenting the factors based on which an ingredient is considered suitable by the model to the user will aid the user to make an informed decision. For example, if a model is trained only to recommend ingredients to users based on carbohydrate content, potatoes might be considered suitable to some users. However, potatoes have a higher glycemic index compared to broccoli (\cite{glycemicindex2024}). While certain diabetic users may be able to handle high glycemic index, some users may not. Hence presenting the list of factors considered to recommend an ingredient as substitute is essential. Currently, the methodologies utilized for ingredient substitution fall into three major categories namely, rule-based approaches, vector embedding-based approaches and knowledge graph-based approaches. Among them, rule-based and knowledge graph-based approaches inherently possess transparency, as their final predictions can be traced back through a series of rules or traversals. The research works that utilize inference on knowledge graph can easily explain the rationale behind the recommendation for ingredient substitution recommendation. However, vector embedding-based approaches present a challenge with respect to transparency. There are several areas of research that focuses on implementing algorithms to explain the outcome of vector embedding-based models, existing works on ingredient substitution models paid little to no attention to it. It is also to be noted that even though the rule-based and knowledge graph-based approaches can be explainable, the explanations are presented to the user and the trustworthiness aspect is yet to be evaluated. To summarize, there is a need for an approach that can scale well by utilizing vector embedding models and can utilize knowledge graphs to provide user-level explanations to ensure the trustworthiness of the models.

\subsubsection{Trustworthiness of the data and knowledge sources}
With AI models, the principle of "garbage in, garbage out" is well established, underscoring the critical importance of data quality. It is essential to ensure that the data and knowledge sources utilized are diverse, comprehensive, and derived from trusted origins. Trusted sources are crucial to maintain the integrity and reliability of the AI model's outputs which is crucial for healthcare domains. Therefore, rigorous validation and curation of data sources are fundamental to developing effective and trustworthy AI systems. Some substitution methods for recipes gather information from websites or user comments on recipe platforms. However, the safety and reliability of these substitutions are debatable since the suggestions are provided by users who are not necessarily domain experts, and the datasets have yet to be validated by professionals in the field. Additionally, some approaches generate substitutions based on the co-occurrence of ingredients. While this method relies on the frequent pairing of certain ingredients, it is not always intuitive or valid. For example, although salt and pepper often appear together in recipes, they cannot be substituted for one another due to their distinct roles in flavoring. Several studies utilize established knowledge sources such as FoodOn, USDA nutrition data (\cite{shirai2020semantics}) and OpenFoodFacts (\cite{loesch2024automated, loesch2022automated}), which are crowdsourced repositories curated by human contributors. However, the community views crowdsourced knowledge with a blend of optimism and caution. On one hand, they are appreciated for their ability to aggregate diverse perspectives and vast amounts of data, which can enhance the richness and scope of AI models. These graphs leverage the collective intelligence of a broad user base, enabling the rapid accumulation and updating of information. However, concerns persist regarding the accuracy, consistency, and potential biases of the data contributed by the crowd, especially in a critical domain such as diet. Knowledge validation by domain experts could aid us in harnessing the full potential of crowdsourced knowledge graphs while mitigating their inherent risks.

\subsubsection{Adherence to Medical Guidelines}
A few existing works present the potential for adaptation of their substitution models for health conditions without referencing adherence to medical guidelines from trustworthy sources such as MayoClinic, WebMD, or CDC. It is also essential to consider all aspects of an ingredient with respect to a particular health condition. For instance, the carbohydrate content of an ingredient is not a sufficient indicator of its suitability for diabetes. Factors such as the type of carbohydrates (simple or complex), the glycemic index of the ingredient, and other nutrients such as fat and protein need to be taken into account as well \footnote{\url{https://glycemicindex.com}}. Such comprehensive consideration is essential to ensure the safety of the model for the user. It is also crucial to evaluate such systems with domain experts to ensure adherence to expected medical guidelines for a given health condition. Certain other works, such as those by \cite{loesch2022automated} and \cite{shirai2021identifying}, seem to recommend “healthy” ingredient substitutions. However, the measure of healthiness is subjective and not clearly defined.

\subsubsection{Summary} With the growth of large models, the safety of AI systems is gaining importance more than ever. These black box models have shown to produce responses that are very believable yet nonsensical or untrue (\cite{ji2023towards, rawte2023troubling}). The issue is commonly referred to as hallucinations. In order to mitigate these issues, several measures towards safe AI are being considered. TrustLLM (\cite{sun2024trustllm}) was such an effort. In certain high stake areas, safe and explainable AI have become mandatory.  It is evident that any model that recommends diet or ingredients to users must adhere to explainable approaches or results, rely on trusted data and knowledge sources and follow medical guidelines. There is a lot that is left to be desired in building safe AI systems for ingredient substitution.

\section{DISCUSSION AND FUTURE DIRECTIONS}
\begin{table}[htbp]

\caption{Summary and potential future possibilities for each research question proposed}
\label{tab:insight}
\begin{longtable}{
  p{0.4\linewidth} p{0.5\linewidth}
}
\hline
\textbf{Factors} & \textbf{Insights or Gap identified}\\
\hline
RQ1: What are the commonly used data sources and techniques for ingredient substitution dataset creation?                                             & The datasets discussed are primarily from recipe websites and user-generated content, including comments and ratings, with some focusing on allergy-safe cooking and regional cuisine styles.\\ \hline                                                                                        
RQ2: What are the technological approaches proposed to address challenges in ingredient substitution?                                                 & Most research relies on vector embeddings for tasks like similarity calculation and pattern recognition. However, recent studies show a shift toward knowledge graph-based approaches, particularly neuro-symbolic methods. These combine knowledge graphs with deep learning to improve accuracy in ingredient substitution and offer insights into the reasoning behind recommendations.\\ \hline
RQ3: What are the additional contextual factors of ingredients that are considered while recommending a substitute ingredient for a given ingredient? & Current research in food ingredient substitution often neglects contextual information such as health labels, disease-related factors, and ingredient flavor profiles. Integrating health labels could help create recipes tailored to dietary needs like vegan, gluten-free, or low-sodium diets. Additionally, the flavor profiles available in databases like FlavorDB are underutilized. Incorporating these elements could significantly improve the precision and relevance of substitution recommendations.\\
\hline
RQ4: What are the dietary and health constraints for which ingredient substitution techniques have been developed?                                    & Ingredient substitution applications serve specific purposes, including vegan-friendly options, allergy-safe alternatives, healthier ingredient recommendations, addressing ingredient unavailability, suggesting local cuisine-based substitutions, and promoting low-carb options. To advance research in these areas, tailored datasets validated by experts like nutritionists are essential to meet specific dietary needs.\\ \hline
RQ5: How safe are these models or systems in terms of transparency and trustworthiness?                                                               & As large AI models grow, concerns about safety have increased, particularly due to hallucinations. Initiatives like TrustLLM are tackling these issues by promoting safe, explainable AI systems. This is crucial in critical areas like dietary recommendations, where reliability and adherence to medical guidelines are essential for safety.\\
\hline
\end{longtable}
\end{table}

Table \ref{tab:insight} provides an overview of insights or gaps identified in each section regarding food ingredient substitution, achieved through a re-examination of the five research questions. The first question examines commonly used data sources and techniques for creating ingredient substitution datasets, highlighting the diversity of datasets sourced from recipe websites and user-generated content. Technological approaches for addressing substitution challenges, including vector embedding- and knowledge graph-based methods, are explored in response to the second question, emphasizing the emerging role of neuro-symbolic computation. The third question addresses additional contextual factors considered in ingredient substitution, underscoring the overlooked aspects of health labels, diseases, and ingredient flavor information. Dietary and health constraints targeted by substitution techniques, such as vegan-friendly or allergy-safe alternatives, are detailed in response to the fourth question, advocating for tailored datasets aligned with specific dietary needs. Finally, the fifth question probes the transparency and trustworthiness of models, highlighting concerns over safety and the need for explainable AI in dietary recommendation systems. Together, these insights and gaps underscore the complexity and evolving nature of research in food ingredient substitution.

\subsection{Future Directions}
\subsubsection{RQ1: Datasets for Ingredient Substitution}
While existing works in food ingredient substitution have produced extensive datasets and evaluation metrics as required for their research, it is limited by the lack of established benchmark repositories. This absence prevents direct comparisons and evaluations of algorithmic performance across different models, making it difficult to assess the effectiveness and robustness of various approaches. Establishing standardized benchmarks would facilitate more meaningful evaluations and advancements in this field. In addition, to improve ingredient substitution datasets, future approaches might need to focus on sourcing data from trusted culinary databases and expert-authored cookbooks rather than solely relying on ingredient co-occurrence. This shift ensures that substitutions are based on functional properties like taste and texture, not mere co-presence in recipes. Social media and recipe website comments can provide practical substitution insights from everyday users. Advanced language models can generate potential pairs by analyzing extensive culinary texts. These generated pairs should be verified by domain experts to ensure trustworthiness and reliability. In addition, adherence to linked data principles facilitates seamless integration with external datasets on the web, expanding the knowledge base, which is useful for ingredient substitutions. To satisfy various use cases, different types of datasets, such as those catering to vegan or dairy-free diets, need to be interconnected and readily retrievable based on structured user queries. 

\subsubsection{RQ2: Technical Approaches}
Vector embeddings and knowledge graph-based approaches are complementary to each other. Vector embeddings are effective for capturing synonyms and excellent for pattern mining and generalization. For instance, if our database contains a recipe for “tomato sauce pasta,” an unknown recipe not in our database, such as “cream sauce pasta,” can easily be understood to belong to the category of pasta or not. While vector embeddings lack semantics as they consist solely of numerical values, knowledge graphs can complement vector embeddings. Knowledge graphs offer semantics by providing the meaning of each node due to their use of natural language. The inherent strength of knowledge graphs lies in their ability to represent semantic relationships through RDF triples, comprising entities, relationships, and attributes. In addition, while the vector embedding approach lacks contextual information, knowledge graphs can provide contextual information, such as flavor, nutrition, and other dietary restrictions based on properties such as glycemic index. Besides capturing domain knowledge, knowledge graphs also facilitate the construction of personalized user knowledge graphs, which in turn facilitates recommending ingredients based on the user’s health condition and food preferences (RQ4). Similarly, the rule-based systems can aid in incorporating rules while inferring the suitability of ingredients as a substitute. For example, grilling meat at high temperature can introduce traces of carcinogens (\cite{salmon1997effects, trafialek2014dietary}). Therefore, the future work on ingredient substitution should focus on neuro-symbolic approaches \cite{sheth2023neurosymbolic} that focuses on bringing learning-based systems, knowledge graphs and rule-based systems together to build an ingredient substitution system. 

\subsubsection{RQ3: Contextual Knowledge of Ingredients} 
Thus far, only nutrition or visual representation of the ingredients are being used as additional contextual knowledge of ingredients. However, ingredient’s flavor, class of ingredients and also suitable cooking methods on ingredients also play a vital role. In addition, if the user has health conditions or dietary preferences, the health labels of the ingredient and disease specific knowledge also plays a crucial role. For example, the health labels can be vegan, lactose free, or gluten free. The disease specific information may include glycemic index of the ingredients for diabetes, or dietary inflammatory index (\cite{foster1995international, shivappa2014designing}) for cardiovascular disease. It is also to be noted that certain disease conditions have implicit constraints that need to be captured through rules such as celiac disease requires a gluten-free diet (\cite{fasano2012celiac}). Such comprehensive contextual knowledge is essential to provide appropriate recommendation of an ingredient as a substitute to another ingredient. Gathering the required contextual knowledge for ingredient substitution based on the end goal is essential to proceed further. 

\subsubsection{RQ4: Constraints for Ingredient Substitution:}
Most existing works have focused on providing ingredient substitutions based on ingredient unavailability, allergen-free, or vegan diets. Several works developed models to evaluate the ingredient substitution model based on the substitution pairs available in the dataset and not based on any specific constraints. As a future direction, research could focus on developing models that address health-related concerns, particularly by leveraging datasets curated with meticulous care by medical field experts. Currently, there is limited research addressing ingredient substitution in the context of health constraints, such as diabetes. In applications involving health constraints, commonly used nutritional information may prove insufficient. More complex considerations, including potential chemical interactions between ingredients, would be necessary to ensure safe and effective substitutions.

\subsubsection{RQ5: Safety of the Ingredient Substitution Models}
As discussed in section 5.5, it is necessary to ensure the safety of the models designed for ingredient substitution. This can be ensured by gathering data from trusted sources such as recipe websites rather than social media posts which may contain incorrect information. Similarly, when suggesting an ingredient for diet constraint (vegan, keto) or diet restriction (lactose-free, gluten-free), we need to ensure that the model adheres to medical guidelines obtained from experts or other medical websites. It is also essential to ensure that the system can explain and reason over the recommendations provided and the reasoning must include the information source. Lastly, the system should be evaluated by necessary domain experts before it is made public to the general population. Periodical evaluation of the system for safety by domain experts can also prevent adverse consequences.

\begin{figure}[!h]
  \begin{center} 
    \includegraphics[width = 1\textwidth]{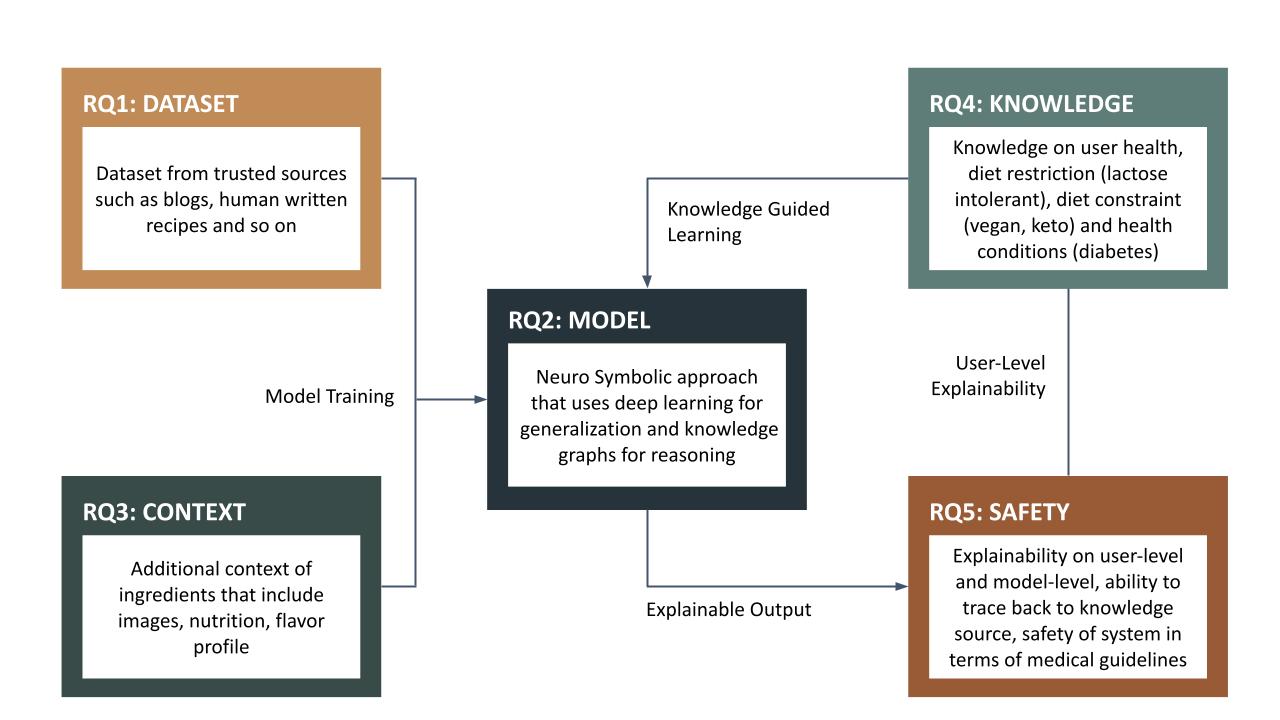}
    \caption{A Neuro-symbolic Framework for Ingredient Substitution that incorporates safety and trustworthiness}
  \label{fig:Neuro-symbolic}
  \end{center} 
\end{figure}

\subsection{Open Challenge}
Food ingredient substitution research lies in developing a comprehensive system that integrates diverse knowledge sources and advanced computational techniques as shown in Figure \ref{fig:Neuro-symbolic}. A critical first step is the creation of rich, reliable datasets sourced from trusted culinary databases, expert-authored cookbooks, and verified user insights from recipe websites (RQ1). These datasets should encompass not only ingredient pairings but also detailed contextual knowledge about ingredients, including their flavors, nutritional profiles, cooking methods, and health-related properties such as glycemic index or dietary inflammatory index (RQ3). Additionally, the system must incorporate disease-specific information and user health knowledge graphs to account for individual dietary needs and restrictions (RQ4).
Building upon this foundation, the challenge calls for the development of a neuro-symbolic model that combines deep learning techniques with knowledge-based approaches (RQ2). This hybrid model would leverage vector embeddings for pattern recognition and generalization, while also utilizing knowledge graphs to provide semantic context and rule-based systems to enforce dietary and health-related constraints. The ultimate goal is to create an intelligent system capable of recommending ingredient substitutions that are not only culinary appropriate but also tailored to individual health needs. Crucially, this system must prioritize safety and transparency, providing clear explanations and reasoning for its recommendations, including citations of information sources (RQ5). It is important to note that currently, there is no comprehensive ingredient substitution knowledge graph that integrates all the resources presented in the supplementary section, highlighting the need for further research and development in this area.

\section{Conclusion}
This survey comprehensively explored the state-of-the-art in food ingredient substitution research, addressing five key research questions. RQ1 examines how various data sources and datasets have been leveraged for ingredient substitution, including recipe websites, user comments, and manually curated datasets from expert guidelines. This section also highlights that the availability of comprehensive benchmark datasets remains limited. RQ2 presents analysis on techniques and approaches used for implementation of ingredient substitution systems. They include a diverse range of techniques spanning rule-based systems, knowledge graphs, ontology design, vector embeddings, and deep learning models. RQ3 discusses external knowledge and additional context that has been leveraged for ingredient substitution task which includes FlavorDB, ingredient images and context of recipes to provide appropriate substitutions. RQ4 studies the conditions and applications of ingredient substitutions in existing literature. Thus far, ingredient substitution systems have been built to recommend vegan diet, avoid allergens, low-carb recommendation, “healthy” ingredient recommendation and for missing ingredients. RQ5 investigates the existing ingredient substitution systems for transparency and trustworthiness. Knowledge validation by domain experts is crucial to mitigate risks and ensure the safety and trustworthiness of AI systems for ingredient substitution. In conclusion, while significant progress has been made in food ingredient substitution research, there remain opportunities for improvement. Future work should focus on curating high-quality datasets, incorporating comprehensive contextual factors, adhering to medical guidelines, validating knowledge sources, and conducting rigorous evaluations involving domain experts and end-users.

\section{Acknowledgments}
We sincerely extend our thanks to Chathurangi Shyalika from University of South Carolina and Kaiping Zheng from National University of Singapore for sharing their expertise in enhancing our paper.

\bibliographystyle{unsrtnat}
\bibliography{references}  
\newpage
\appendix
\section*{Appendices}
\section{List of Recipe and Ingredient Websites}
\label{appendix:A1}
\begin{itemize}
\item AjinomotoCo., Inc.: \url{http://www.ajinomoto.co.jp/recipe/}
\item Allerhande: \url{http://www.ah.nl/allerhande/recept}
\item AllRecipes: \url{https://www.allrecipes.com/}
\item BBC Food Recipe: \url{https://www.bbc.co.uk/food/recipes}
\item Chefkoch: \url{https://www.chefkoch.de/}
\item Cookpad: \url{https://cookpad.com/us}
\item CookStr: \url{https://www.cookstr.com/}
\item Epicurious: \url{https://www.epicurious.com/}
\item Food.: \url{http://www.food.com/recipes#}
\item FoodSubs: \url{http://www.foodsubs.com/}
\item Kochbar: \url{https://www.kochbar.de/}
\item Kyounoryouri: \url{http://www.kyounoryouri.jp/}
\item Spoonacular: \url{https://www.spoonacular.com/}
\end{itemize}

\section{List of Data Sets}
\begin{itemize}
\item French dataset INCA 21: \url{https://www.data.gouv.fr/fr/datasets/donnees-de-consommations-et-habitudesalimentaires-de-letude-inca-2-3/}
\item Loesh et al, 2022: \url{https://www.kaggle.com/shuyangli94/food-com-recipes-and-userinteractions}
\item FoodOn: \url{https://foodon.org/}
\end{itemize}


\end{document}